\documentclass[12pt,axodraw]{article}

\usepackage{axodraw}

\def\bra{\langle}
\def\ket{\rangle}

\def\beq{\begin{equation}}
\def\eeq{\end{equation}}
\def\bea{\begin{eqnarray}}
\def\eea{\end{eqnarray}}

\begin{document}

Unitary symmetry of sum rules for hadron photoproduction

on octet baryons

V S. Zamiralov

Comments: 7 pages, 1 figure, uses axodraw.sty

\footnote{zamir@depni.sinp.msu.ru}

Skobeltsyn Institute of Nuclear Physics,
Lomonosov Moscow State University,Moscow, Russia 

Dubnicka-Dubnickova-Kuraev (DDK) sum rules are considered.
It is shown that integrals over differences
of the total photoproduction cross-sections on octet baryons
could be understood in terms of unitary symmetry approach.
All the DDK sum rules for these quantities are expressed in
terms of only three parameters.

High Energy Physics - Phenomenology

\section{Introduction}

Many years ago Gottfried \cite{Gottfried} wrote dispersion sum rule
interlacing proton magnetic moment and proton charge radius
with integral over total photoproduction cross-section on protons:
\beq
\int^{\infty}_{0}
\frac{d\nu}{\nu}[\sigma_{tot}^{\gamma B\rightarrow X}(\nu)=
4\pi^2\alpha[\frac13\bra r^2_{Ep}\ket +\frac{1-\mu^2_p}{4m^2_p}]
\eeq
where $\bra r^2_{Ep}\ket$ is the proton mean square charge radius
and $\mu_p=1+\kappa_p$ and $\kappa_p$ are total and anomalous
magnetic moments of proton in terms of nuclear magnetons.

As  years ago it was proved experimentally
that total photoproduction cross-section on protons is
arising with energy \cite{PDG} the Gottfried
sum rule results to contain  diverging integral and
therefore cannot be valid.

But recently an interesting sum rules for photoproduction
of baryons were proposed which overcome this difficulty
considering instead  differences between
integrals over total cross sections on various baryons \cite{DDK1}.
In this way convergency of the integral was achieved
and series of sum rules were evaluated
\cite{DDK, BDDK}. The main assumption for the convergency lies in
equality of the Pomeron exchange for all the baryons
of the octet. It is seems to be valid for the baryons of the same
isomultiplet and plausible for the whole unitary octet.

\section{DDK universal sum rules}

The DDK universal sum rule \cite{DDK}
can be written as the equality with left-hand side (LHS) in terms of baryon anomalous
magnetic moments and Dirac baryon mean square radii  and
right-hand side (RHS)
as integral over difference of total photoproduction cross-sections
on octet baryons
\bea
\frac13[F_{1B}(0)\bra r^2_{1B}\ket -F_{1B^\prime}(0)\bra r^2_{1B^\prime}\ket]
-[\frac{\kappa_B^2}{4m_B^2} -\frac{\kappa_{B^\prime}^2}{4m_{B^\prime}^2}]=
\eea
$$
\frac{2}{\pi^2\alpha}\int^{\infty}_{\omega_B}
\frac{d\omega}{\omega}[\sigma_{tot}^{\gamma B\rightarrow X}(\omega)-
\sigma_{tot}^{\gamma {B^\prime}\rightarrow X}(\omega)],
$$
which relates Dirac baryon mean-square radii $\bra r^2_{1B}\ket$,
$\bra r^2_{1B^\prime}\ket$
 and anomalous magnetic moments  $\kappa_B$ , $\kappa_B^\prime$
to the convergent integral due to presumed cancellation
of the otherwise arising  high energy total cross-sections at
$\omega\rightarrow\infty$. Electric form factors
$F_{1B}(q^2)$, $F_{1B^\prime}(q^2)$ reduce to electric
charges $e_B$, $e_{B^\prime}$ at zero momentum transfer squared $q^2=0$.
Instead the Dirac baryon mean-square radii $\bra r^2_{1B}\ket$ can be
reliably taken from the
relation
\beq
\bra r^2_{EB}\ket=\bra r^2_{1B}\ket+3\frac{\kappa_B}{4m_B^2}
\eeq
either from experimental data (for $p$, $n$ and $\Sigma^-$)\cite{PDG}
or from theory \cite{Kubis}.
We put them into the Table 1. ( Note that sum of the  4th and 5th
coluns just give the 3rd one. )

We remind the first sum rule of \cite{DDK1}
\bea
\frac13 \bra r^2_{1p}\ket-\frac{\kappa_p^2}{4m_p^2}+
\frac{\kappa_n^2}{4m_n^2}=
\frac{2}{\pi^2\alpha}\int^{\infty}_{\omega_N}
\frac{d\omega}{\omega}[\sigma_{tot}^{\gamma p\rightarrow X}(\omega)-
\sigma_{tot}^{\gamma n\rightarrow X}(\omega)]
\label{DDK}
\eea
with $\omega_N=m_\pi+m_\pi^2/2M_N$.

Here $\bra r^2_{1p}\ket$ is electric square radus of the proton
and  $\kappa_{p,n}$ mean  anomalous magnetic moments of proton and
neutron. It agrees well with experiment as
LHS=1.93$\pm$ 0.18 mb and RHS=1.92$\pm$ 0.32 mb \cite{DDK1}.

It was generalized in \cite{DDK} to all the octet baryons
writing 28 relations. But as it is easy to see only 7 of them are
linearly independent, and we choose them as Eq.(1) plus other 6 relations
below.
The important issue is that while treating photoproduction on
$\Lambda$ ( or $\Sigma^0$) we should also add the $\Sigma^0\Lambda$
contribution. For the differencies $\Sigma^0-\Lambda$
 these contributions cancel each other.
But in other cases with single $\Sigma^0$ or single  $\Lambda$
in pair with any other baryon we have found noticeable effects.

$$
\frac13 \bra r^2_{1\Sigma^+}\ket-\frac{\kappa_{\Sigma^+}^2}{4m_{\Sigma}^2}+
\frac{\kappa_{\Sigma^0}^2}{4m_{\Sigma}^2}+
\frac{\kappa_{\Sigma^0\Lambda}^2}{4m_{\Sigma\Lambda}^2}=
\frac{2}{\pi^2\alpha}\int^{\infty}_{\omega_N}
\frac{d\omega}{\omega}[\sigma_{tot}^{\gamma {\Sigma^+}\rightarrow X}(\omega)-
\sigma_{tot}^{\gamma \Sigma^0\rightarrow X}(\omega)];
$$

$$
\frac13 \bra r^2_{1\Sigma^-}\ket-
\frac{\kappa_{\Sigma^0}^2}{4m_{\Sigma}^2}-
\frac{\kappa_{\Sigma^0\Lambda}^2}{4m_{\Sigma\Lambda}^2}+
\frac{\kappa_{\Sigma^-}^2}{4m_{\Sigma}^2}=
\frac{2}{\pi^2\alpha}\int^{\infty}_{\omega_N}
\frac{d\omega}{\omega}[\sigma_{tot}^{\gamma \Sigma^0\frac{\kappa_{\Sigma^0\Lambda}^2}{4m_{\Sigma\Lambda}^2}\rightarrow X}(\omega)-
\sigma_{tot}^{\gamma \Sigma^-\rightarrow X}(\omega)]
$$

\bea
\frac13 \bra r^2_{1\Sigma^+}\ket-\frac{\kappa_\Lambda^2}{4m_\Lambda^2}+
\frac{\kappa_{\Sigma^+}^2}{4m_\Sigma^2}+
\frac{\kappa_{\Sigma^0\Lambda}^2}{4m_{\Sigma\Lambda}^2}=
\frac{2}{\pi^2\alpha}\int^{\infty}_{\omega_N}
\frac{d\omega}{\omega}[\sigma_{tot}^{\gamma \Lambda\rightarrow X}(\omega)-
\sigma_{tot}^{\gamma \Sigma^+\rightarrow X}(\omega)]
\label{DDK1}
\eea
$$
\frac13 \bra r^2_{1\Xi^-}\ket-\frac{\kappa_{\Xi^0}^2}{4m_\Xi^2}+
\frac{\kappa_{\Xi^-}^2}{4m_\Xi^2}=
\frac{2}{\pi^2\alpha}\int^{\infty}_{\omega_N}
\frac{d\omega}{\omega}[\sigma_{tot}^{\gamma \Xi^0\rightarrow X}(\omega)-
\sigma_{tot}^{\gamma \Xi^-\rightarrow X}(\omega)]
$$
$$
\frac13 \bra r^2_{1p}\ket-
\frac13 \bra r^2_{1\Sigma^+}\ket-\frac{\kappa_p^2}{4m_p^2}+
\frac{\kappa_{\Sigma^+}^2}{4m_\Sigma^2}=
\frac{2}{\pi^2\alpha}\int^{\infty}_{\omega_N}
\frac{d\omega}{\omega}[\sigma_{tot}^{\gamma p\rightarrow X}(\omega)-
\sigma_{tot}^{\gamma \Sigma^+\rightarrow X}(\omega)]
$$
$$
\frac13 \bra r^2_{1p}\ket-\frac{\kappa_p^2}{4m_p^2}+
\frac{\kappa_{\Xi^0}^2}{4m_n^2}= 
\frac{2}{\pi^2\alpha}\int^{\infty}_{\omega_N}
\frac{d\omega}{\omega}[\sigma_{tot}^{\gamma p\rightarrow X}(\omega)-
\sigma_{tot}^{\gamma \Xi^0\rightarrow X}(\omega)]
$$

We control relations by
putting also two corollaries:
$$
\frac{2}{\pi^2\alpha}\int^{\infty}_{\omega_N}
\frac{d\omega}{\omega}[\sigma_{tot}^{\gamma \Sigma^+\rightarrow X}(\omega)-
\sigma_{tot}^{\gamma \Sigma^-\rightarrow X}(\omega)]
$$
=4.0131 mb (4.2654 mb \cite{DDK})

$$
\frac{2}{\pi^2\alpha}\int^{\infty}_{\omega_N}
\frac{d\omega}{\omega}[\sigma_{tot}^{\gamma n\rightarrow X}(\omega)-
\sigma_{tot}^{\gamma \Xi^0\rightarrow X}(\omega)]
$$
= -0.4075+0.0875=-0.3200 mb (-0.3156 mb \cite{DDK})
\begin{center}
\begin{picture}(200,80)(0,0)
\ArrowLine(0,40)(40,40)
\ArrowLine(40,40)(70,70)
\Photon(40,40)(70,20)3 4
\ArrowLine(70,20)(110,40)
\ArrowLine(70,20)(100,10)
\ArrowLine(0,20)(70,20)
\Vertex(70,20)3
\Text(70,40)[c]{$ \gamma $}
\Text(0,45)[c]{$ e^{-} $}
\Text(70,75)[c]{$ e^- $}
\Text(100,25)[l]{$ X $}
\Text(10,30)[l]{$ B $}
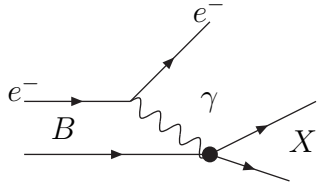
\Text(80,0)[c]{ Fig.1. Inclusive photoproduction on
octet baryon B 1/2$^+$  }
\end{picture}
\end{center}

\section{Unitary symmetry relations}

We study now unitary symmetry of these sum rules for the
differences of the integrals of total photoproduction
cross-sections on baryons of the octet 1/2$^+$. As we will see all of them
could be described in terms of only 3 parameters.

We remind universal formula for magnetic moments which transfers either
in the NRQM formula (with $F$=2/3 and $D$=1 and $e_q\rightarrow \mu_q$)
or into the unitary symmetry result (by putting
quark electic charge explicitly) \cite{Gelmi}:
\beq
\mu_{\Sigma^0}=(e_u+e_d)F+e_s(F-D).
\eeq
All the other baryon magnetic moments but that of $\Lambda$
are obtained just by changing properly quark indices.
Instead that of the $\Lambda$ baryon could be written as
\beq
\mu_{\Lambda}=\frac13[2\mu_{\Sigma^{ds}}+2\mu_{\Sigma^{us}}-
\mu_{\Sigma^0}]=
\frac13[(e_u+e_d+4e_s)F+(2e_u+2e_d-e_s)](F-D)].
\eeq
For the corresponding transition moment one get
\beq
\sqrt3\mu_{\Sigma^0\Lambda}=\mu_{\Sigma^{ds}}-\mu_{\Sigma^{us}}
=(e_u-e_d)D,
\eeq                        
where subscribes $ds$ and $us$ mean just that we interchanged
quarks $d\leftrightarrow s$ and $u\leftrightarrow s$ 
in the $\Sigma^0$ wave function.

We have seen that similar reasoning can be applied to the QCD sum rules
not only for magnetic moments \cite{Oz} but also for other
vertex quantities \cite{Aliev}. But in the QCD sum rules
the quantities analogues to $F$, $D$ would contain the dependence on the quark
parameters absent in the simple unitary symmetry model \cite{Oz}.

We now proceed with differences of the total cross-sections on
octet baryons by using as operator not the electric charge of quarks
but
their electric charge squared. That is, we describe
finite parts of the integrals in Eqs.(\ref{DDK},\ref{DDK1}) 
which we denote just by
symbol of the target, in terms of $\cal{F,E}$'s depending on the quark
parameters  at this stage only phenomenologically putting subindices
$s$ and $ss$ to indicate number of strange quarks in baryon 
\beq
{\cal{P}}={2e_u^2 }{\cal{F}}+{e_d^2}{\cal{E}}
\eeq
\beq
\Sigma^0={(e_u^2+e_d^2)} {\cal{F}}^s+e_s^2{\cal{E}}^s
\eeq
\beq
\Xi^0={2e_s^2 }{\cal{F}}^{ss}+{e_s^2}{\cal{E}}^{ss}
\eeq
and so on.

Upon using our relations \cite{Gelmi, Oz}
we write for the finite $\Lambda$ contribution
\beq
\Lambda=(1/3)[{(e_u^2+e_d^2+4e_s^2)} {\cal{F}}^s+
[2e_u^2+2e_d^2-e_s^2]{\cal{E}}^s]
\eeq
with 
${\cal{E}}^s$=${\cal{E}}$= ${\cal{F}}-{\cal{D}}$ in the unitary limit.

The structure ${\cal{F}}$ corresponds to the contributons
of two (quasi)similar quarks ($uu$,$dd$,$ud$, $ss$), while
the structure ${\cal{E}}$ corresponds to single-quark
contribution \cite{Gelmi,Oz}. Subindices $s$ and $ss$ indicate
number of strange quarks in baryon.

\begin{table}
\begin{center}
\begin{tabular}{|c|c|c|c|c|c|} \hline
B & $\kappa_B[\mu_N]$ & $\bra r^2_{EB}\ket $ [mb]
& $3\kappa_B/2m^2_B $ [mb]  
& $\bra r^2_{1B}\ket$ [mb] &  $\kappa^2_B/4m^2_B $ [mb]\\
\hline
p & 1.7928 & 7.17 & 1.19 & 5.98 & 0.3560\\
\hline
n & -1.9130 & -1.13 & -1.27 & 0.14 & 0.4075\\
\hline
$\Lambda$ & -0.6130 & 1.10 & -0.29 & 1.39 & 0.0295\\
\hline
$\Sigma^+$ & 1.4580 & 6.00 & 0.60 & 5.40 & 0.1458\\
\hline
$\Sigma^0$ & 0.6490 & -0.30 & 0.27 & -0.57 & 0.0293\\
\hline
$\Sigma^-$ & -0.1600 & 6.70 & -0.07 & 6.77 & 0.0019 \\
\hline
$\Xi^0$ & -1.250 & 1.30 & -0.42 & 1.72 & 0.0875\\
\hline
$\Xi^-$ & 0.3493 & 4.90 & 0.12 & 4.78 & 0.0070\\
\hline
\end{tabular}
\caption{ Contributions of magnetic moments and charge radii of  octet
baryons 1/2$^+$ in mb}
\end{center}
\label{table1}
\end{table}

\begin{table}
\begin{center}
\begin{tabular}{|c|c|c|c|c|} \hline
B-B$\prime$ & LHS (mb) & RHS  & RHS  (mb)
& Unitary formulae \\
targets &This Work&Eq.(13)(mb)&DDK[4]&\\
\hline
p-n & 2.0445 &  2.0  & 2.0414 & $(2/3F-1/3E)$\\
\hline
p-$\Sigma^+$ & -0.0172 & 0.0 & -0.4158 & $8/9(F-F^s)+1/9(E-E^s)$\\
\hline
$\Sigma^+$-$\Sigma^0$ & 1.9710 & 2.0 & 2.0825 &
$(1/3F^s)$\\
\hline
$\Sigma^0$-$\Sigma^-$ & 2.0411 & 2.0 & 2.1829 &
$(1/3F^s)$\\
\hline
$\Sigma^0$-$\Lambda$ & 0.0002 & 0.0 & -0.0006 & 0 \\
\hline
$\Xi^0$-$\Xi^-$ & 1.4955 & 1.5 & 1.5921 & $(1/3E^{ss})$\\
\hline
$\Sigma^-$-$\Xi^-$ & -0.658 & -0.66 & -0.5921 &
$2/9(F^s-F^{ss})+1/9(E^s-E^{ss})$\\
\hline
\end{tabular}
\caption{ DDK sum rules for baryons 1/2$^+$.
In the 3rd column $F=F^s=E=E^s=6.0, F^{ss}=9.72, E^{ss}=4.5$ are used.}
\end{center}
\label{table2}
\end{table}
Thus we can write the righthand sides of the DDK sum rules,
that is, the integral over differences of the total cross-sections
on two different baryons $B$ and $B^\prime$.
We choose $p$ and $\Sigma^+$ putting the rest into the Table 2 and
put them into the form
\beq
[2e_u^2{\cal F}+e_d^2{\cal E}]-[2e_u^2{\cal F}^s+e_s^2{\cal E}^s]=
\int^{\infty}_{\omega_{thresh}}
\frac{d\omega}{\omega}
(\sigma^{\gamma p\rightarrow X }(\omega)-
\sigma^{\gamma\Sigma^+\rightarrow X }(\omega)).
\eeq
We repeat that while treating photoproduction on
$\Lambda$ ( or $\Sigma^0$) we should also add the $\Sigma^0\Lambda$
contribution. For the differencies $\Sigma^0-\Lambda$
these contributions cancel each other.
But in other cases with single $\Sigma^0$ or single  $\Lambda$
in pair with any other baryon we have obtained noticeable effects.

We have succeeded to fit
the RHS's of Eqs(\ref{DDK},\ref{DDK1}) with $F=E= F^s=E^s=6$ and
$F^{ss}=3.3, E^{ss}=4.5$ (see Table 2).
Only cascade hyperon contributions differ strongly
from unitary symmetry parameters which can be explained by
presence of two heavy quarks and only one light quark in these hyperons.
Discovery of doubly charmed baryons $\Xi_{cc}$ \cite{PDG} can help us to
solve this disaccord.

\section{Conclusion}

We have shown that unitary symmetry of the BDDK sum rules for the
differences of the integrals of total photoproduction
cross-sections on baryons of the octet 1/2$^+$ holds
within some reasonable lines. In fact we succeeded
in describing DDK sum rules in terms of only 3 parameters.
It is also shown that one should take into account not only
$\Sigma^0$ and $\Lambda$ baryons  while analysing total
cross-sections $\sigma^{\gamma\Sigma^0\rightarrow X }$ and
$\sigma^{\gamma\Lambda\rightarrow X }$
but also take into account
the $\Sigma^0\Lambda$ transition mode.

I am grateful to Takhmassib Aliev and Boris Ishkanov for
valuable discussions. I have also in memory a
conversation years ago on the subject with Eduard Kuraev who
prematurely has left us in 2014.


\newpage
\begin{thebibliography}{99}
\bibitem{Gottfried} K.Gottfried, Phys. Rev. Lett. {\bf 18}(1967)1174.
\bibitem{PDG}C.Patrignani {\it et al.}(PDG)  Chin. Phys. C40, 100001
(2016) and 2017 update.
\bibitem{DDK1} S.Dubnicka, A.Z.Dubnickova and E.A.Kuraev,
Phys. Rev. D {\bf 73} (2006) 034023.
\bibitem{DDK} S.Dubnicka, A.Z.Dubnickova and E.A.Kuraev,
Phys. Rev. D {\bf 75} (2007) 057901;
\bibitem{BDDK}
 E.Bartos, .Dubnicka, A.Z.Dubnickova and E.A.Kuraev,
FIZIKA B (Zagreb) {\bf 17} (2008) {\bf 1}, 11-24.
\bibitem{Kubis}B.Kubis and U.-G.Meissner, Eur. Phys. J. C {\bf 18}
(2001) 747.
\bibitem{Aliev80}T.M.Aliev,A.Ozpineci,M.Savci, and V.Zamiralov,
Phys.Rev.D80,01610105 (2009)
%
\bibitem{Aliev}T.M.Aliev,A.Ozpineci and V.Zamiralov, Journ.Phys.:
conf. Series {\bf 348} (2012) 012009
\bibitem{Gelmi}L.Gelmi, V.S.Zamiralov, and S.N.Lepshokov,
MSU Bulletin. Ser.3. Phys. Astron. {\bf 2}, 33 (1989).
\bibitem{Oz}A.Ozpineci,S.Yakovlev, and V.Zamiralov, 
At.Nucl.Phys.{\bf 68},304(2005);
Mod.Rev.Lett.{\bf 20}, 243(2005)
\end{thebibliography}
\end{document}